\documentclass[aps,twocolumn,groupedaddress]{revtex4}
\usepackage[dvips]{color}
\usepackage{amssymb}

\newcommand{\ped}[1]{\ensuremath{_{\rm #1}}}

\definecolor{awesome}{rgb}{1.0, 0.13, 0.32}
\usepackage[dvips]{graphicx}
\usepackage{color}
\definecolor{blue}{rgb}{0,0,0}
\begin{document}
\title[Point-contact Andreev-reflection spectroscopy in Fe(Te,Se) films]{Point-contact Andreev-reflection spectroscopy in Fe(Te,Se) films: multiband superconductivity and electron-boson coupling}
\author{D Daghero, P Pecchio, G A Ummarino, R S Gonnelli}
\affiliation{Dipartimento di Scienza Applicata e Tecnologia, Politecnico di Torino, 10129 Italy}
\author{F Nabeshima, Y Imai, A Maeda}
\affiliation{Department of Basic Science, The University of Tokyo, Tokyo 153-8902, Japan}
\author{I Tsukada, S Komiya}
\affiliation{Central Research Institute of Electric Power Industry, Kanagawa 240-0196, Japan}

\begin{abstract}
We report on a study of the superconducting order parameter in $\mathrm{Fe(Te_{1-x}Se_{x})}$ thin films (with different Se contents: $x=0.3, 0.4, 0.5$) by means of point-contact Andreev-reflection spectroscopy (PCARS). The PCARS spectra show reproducible evidence of multiple structures, namely two clear conductance maxima associated to a superconducting gap of amplitude $\Delta_E\simeq 2.75 k\ped{B}T\ped{c}$ and additional shoulders at higher energy that, as we show, are the signature of the strong interaction of charge carriers with a bosonic mode whose characteristic energy coincides with the spin-resonance energy. The details of some PCARS spectra at low energy suggest the presence of a \emph{smaller} and not easily discernible gap of amplitude $\Delta_H\simeq 1.75 k\ped{B}T\ped{c}$. The existence of this gap and its amplitude are confirmed by PCARS measurements in $\mathrm{Fe(Te_{1-x}Se_{x})}$ single crystals. The values of the two gaps $\Delta_E$ and $\Delta_H$, once plotted as a function of the local critical temperature $T\ped{c}^A$, turn out to be in perfect agreement with the results obtained by various experimental techniques reported in literature.
\end{abstract}
\maketitle

\section{Introduction}
The iron-based superconductor $\mathrm{Fe(Te_{1-x}Se_x)}$ has recently been the subject of an intense research effort, both experimental and theoretical. Despite its simpler structure if compared to 122 or 1111 compounds \cite{paglione10}, this material has been challenging the researchers and several of its properties have been (or still are) controversial. From the experimental point of view, for example, the presence of excess Fe has effects on the critical temperature, on the transport properties and on the magnetic properties (i.e. the spin fluctuations) which are not so easy to disentangle from the effects of the Se substitution \cite{nakayama10}. Moreover, one of the fundamental steps in the study of new superconductors is the determination of the number, the symmetry and the amplitude of the superconducting energy gaps, but also in this respect  $\mathrm{Fe(Te_{1-x}Se_x)}$ has long been a puzzle. Various experimental techniques have given conflicting results, ranging from single \cite{nakayama10,park10,kato09,wu10,hanaguri10} or multiple isotropic gaps \cite{homes10,bendele10,miao12,yin14}, to highly anisotropic or nodal gap(s) \cite{zeng10,kim10}. The amplitudes of the gaps and of the gap ratios $2\Delta/k_BT_c$ are considerably scattered as well, so that it is difficult to extract a consistent picture. This is complicated by the fact that all the measurements reported in literature have been performed on a \emph{single} doping content.
In this paper we report on the extensive investigation of the energy gaps of $\mathrm{Fe(Te_{1-x}Se_x)}$ thin films with different Se contents ($x=0.3$, $0.4$ and $0.5$) by means of point-contact Andreev reflection spectroscopy (PCARS). This study provides a consistent picture of how the gaps evolve with the critical temperature, in which most of the data reported in literature perfectly fit. We show that the PCARS spectra actually give evidence of \emph{two} energy gaps, plus a third energy scale which we identify as being due to the strong electron-boson interaction (EBI). The energy $\Omega_0$ of the mediating boson is consistent with the empirical law $\Omega_0 = 4.65 k_B T_c$ that relates the spin resonance energy observed at the wave vector $\mathbf{Q}=(\frac{1}{2}, 0)$ \cite{qiu09} to the critical temperature $T_c$ \cite{paglione10}.

\section{Experimental details}
\begin{figure}
\begin{center}
\includegraphics[width=0.8\columnwidth]{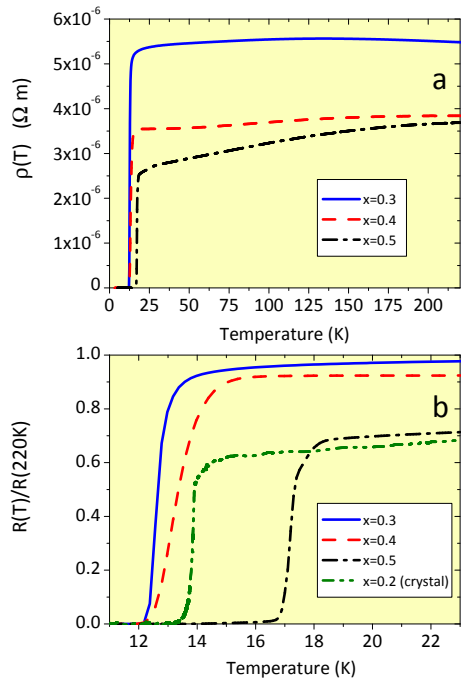}
\end{center}
\vspace{-1cm}
\caption{{\color{blue}{(a) Resistivity vs. temperature for the three films at different Se contents used in this paper}}. The critical temperatures as determined from the resistive transitions are the following: for $x=0.5$, $T\ped{c}^{90}=17.3$ K and $T\ped{c}^{10}=16.8$ K; for $x=0.4$, $T\ped{c}^{90}=14.3$ K and $T\ped{c}^{10}=12.6$ K; for $x=0.3$, $T\ped{c}^{90}=13.3$ K and $T\ped{c}^{10}=12.3$ K. {\color{blue}{(b) Normalized resistance $R(T)/R(220 \, \mathrm{K})$ of the same films and of the $x=0.2$ single crystal used for PCARS measurements, in the region of the superconducting transition. In the case of the single crystal, $T\ped{c}^{90}=14.1$ K and $T\ped{c}^{10}= 13.6$ K.}} }\label{fig:resistivity}
\end{figure}

The films were grown by pulsed laser deposition {\color{blue}{using a KrF excimer
laser (wavelength 248 nm)}} starting from a $\mathrm{Fe(Te_{1-x}Se_{x})}$ target, on top of commercially available $\mathrm{CaF_2}$ (100) substrates . The parameters of the deposition were the following: laser repetition rate 10 Hz, {\color{blue}{laser energy 300 mJ}}, substrate temperature 280$^{\circ}$C, back pressure $10^{-7}$ Torr {\color{blue}{(further details can be found in}} \cite{imai10,tsukada11,ichinose13}). Some of the films were deposited by using a specially designed metal mask, directly put on top of the substrate, in order to obtain a six-terminal shape convenient for transport (resistivity) measurements. Others were instead deposited on the whole substrate (about $0.5 \times 0.5 \, \mathrm{cm^2}$) just to provide a larger area for point-contact spectroscopy measurements. {\color{blue}{In this case the resistivity was measured by using the four-probe van der Pauw configuration.}}
The CaF$_2$ substrate is known to cause a sizable lattice strain in films of $\mathrm{Ba(Fe,Co)_2As_2}$ that results in a stretch of the phase diagram with respect to that of single crystals \cite{pecchio13}. In the case of our $\mathrm{Fe(Te_{1-x}Se_{x})}$ films, the
$a$-axis ($c$-axis) parameter is shorter (larger) than in single crystals \cite{sales09,tsukada11} and the critical temperature is considerably enhanced as well \cite{tsukada11,nabeshima13}.  This effect was recently found to arise mainly from the chemical substitution of anions at the  $\mathrm{Fe(Te_{1-x}Se_{x})/CaF_2}$ interface, while the lattice mismatch plays a secondary role \cite{ichinose13}. However, in the films used for the present PCARS measurements chemical interdiffusion occurs only in a thin region of the order of 10 nm across the interface (see e.g. films A and B of ref.\cite{ichinose13}).  {\color{blue}{Since the thickness of our films is 100 nm for $x$=0.3 and 0.4, and 86 nm for $x$=0.5,}} this does not affect the composition of the films at the surface, as also clearly shown by EDX measurements at different depths \cite{ichinose13}. The same results, as well as the absence of anomalous Hall effect \cite{tsukada10} indicate that the excess Fe is negligible in these films.
As witnessed by XRD measurements, the films do not present detectable amounts of impurity phases, are $c$-axis oriented, and grow with a 45$^{\circ}$ rotation with respect to the underlying CaF$_2$ substrate, i.e. $\mathrm{Fe(Te_{1-x}Se_{x})}\, [100] \parallel \mathrm{CaF_2 }\,[110]$. {\color{blue}{Figure \ref{fig:resistivity}a reports the resistivity of the three films with $x$=0.3, 0.4 and 0.5 as a function of temperature}}. The critical temperatures $T\ped{c}^{90}$ and $T\ped{c}^{10}$, (defined as the temperatures at which the resistivity drops to 90\% and 10\% of its normal-state value just before the superconducting transition) are listed in the caption of the same figure. {\color{blue}{A magnification of the (normalized) curves in the region of the transition is reported in Fig. \ref{fig:resistivity}b}}.

The Fe(Se$_{0.2}$Te$_{0.8}$) single crystals (used here as a term of comparison for films) were grown, as described in ref. \cite{komiya13}, by the Bridgman method starting from stoichiometrically weighed Fe (99.99\% pure), Se (99.999\% pure) and Te (99.999\% pure). The as-grown crystals were then annealed in a moderate vacuum atmosphere ($\simeq 1$ Pa). Unlike annealing in high vacuum, this process has been shown to increase significantly the critical temperature, probably thanks to the formation of an iron oxide layer on the surface (due to the reaction with the residual oxygen) that drags the excess Fe out of the bulk. The oxide layer is then removed mechanically. The crystal composition was determined by microanalysis \cite{komiya13} and turned out to be in good agreement with the nominal one. {\color{blue}{The normalized resistivity of a Fe(Se$_{0.2}$Te$_{0.8}$) single crystal is reported for comparison in Fig. \ref{fig:resistivity}(b)}}.

Point-contact Andreev-reflection spectroscopy (PCARS) measurements were performed by using a ``soft'' pressureless technique in which a thin Au wire ($\varnothing=18 \,\mu\mathrm{m}$) is kept in contact with the sample surface by means of a small drop ($\varnothing \leq 100 \,\mu\mathrm{m}$) of Ag conducting paste. In this way, parallel nanometric contacts (that can well fulfill the requirement for the ballistic or diffusive conduction that are indispensable for PCARS \cite{naidyuklibro}) are established here and there within the area covered by the Ag paste: This means that each PCARS spectrum is actually the result of a spatial average over a microscopic area of the sample surface. The reason to prefer the ``soft'' pressureless technique to the conventional  ``needle-anvil'' one, in which a sharp metallic tip is pressed against the sample surface, is due both to the fragility of the CaF$_2$ substrate and to the much better thermal and mechanical stability of these ``soft'' contacts \cite{gonnelli02c}. Thanks to the $c$-axis orientation of the films, the normal/superconductor interface is always parallel to the $ab$ plane, that means that the probe current is always injected (mainly) along the $c$ axis.
All the contacts were in the regime of Andreev reflection, in which the potential barrier at the N-S interface is low enough to make Andreev reflection dominate over quasiparticle tunneling in the conduction through the contact.

\section{Results and discussion}
\begin{figure}
\begin{center}
\includegraphics[width=\columnwidth]{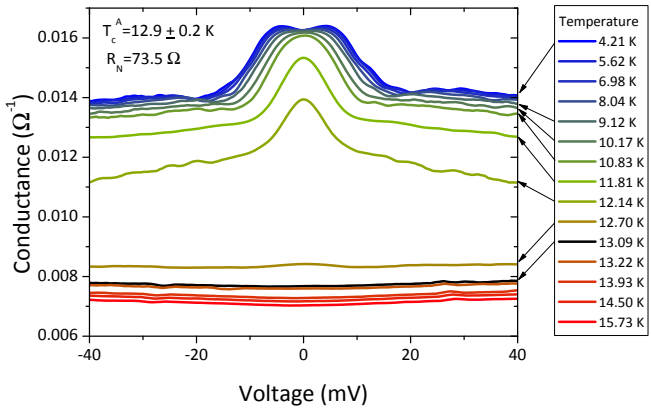}
\end{center}
\vspace{-5mm}\caption{Temperature dependence of the raw conductance curves of a point contact on the film with $x=0.3$. The black curve ($T=13.09$ K) is the normal-state conductance.}\label{fig:Tdep}
\end{figure}
Figure \ref{fig:Tdep} shows the temperature dependence of the differential conductance curve ($dI/dV$ vs. $V$) of a point contact whose resistance is $73.5 \,\Omega$, made on the film with $x=0.3$. The low-temperature curve presents no traces of anomalous features (zero-bias anomalies, downward bending of the high-energy tails, dips and so on) that could suggest the breakdown of the conditions for energy-resolved spectroscopy \cite{daghero10}. Instead, it presents the typical conductance enhancement (with symmetric maxima) caused by Andreev reflection at the interface, superimposed to an almost flat background. On increasing the temperature the Andreev reflection structure is progressively suppressed -- as a consequence of both the thermal smearing and the decrease in gap amplitude -- while the curves shift downward. At $T\geq 13.09$ K the shape of the conductance curves does not change any longer; we identify the relevant conductance curve (black) with the normal-state conductance of the junction. Note that the local critical temperature (that we will call $T\ped{c}^{A}$ from now on) falls somewhere between 12.70 K and 13.09 K and thus we assume conservatively $T\ped{c}^{A}=12.9 \pm 0.2$ K, which lies between $T\ped{c}^{10}$ and $T\ped{c}^{90}$ for this film (see  figure \ref{fig:resistivity}). The downward shift of the conductance curves is typical of films and is due to the spreading resistance contribution arising from the portion of the film between the point contact and the second voltage electrode \cite{chen10b,doring14}. This spreading resistance is zero as long as the temperature is sufficiently low, but starts playing a role when the temperature approaches the critical one; in particular, it gives rise not only to a shift of the curves, but also to a stretching of their horizontal scale \cite{chen10b} and, in the temperature region where the $\rho(T)$ curve is steeper, to an enhanced bending of the PCARS spectra (see the curve at 12.14 K in figure \ref{fig:Tdep}). This means that: i) only the low-temperature curves have the correct voltage scale and can be used for spectroscopic purposes; ii) the conductance curve in the normal state cannot be used for the normalization of these curves, which is necessary in order to compare them with theoretical models and extract the gap values. The normalization therefore, as usual in Fe-based compounds \cite{daghero11}, requires some caution and becomes somewhat arbitrary. We solved the problem by using different (reasonable) normalization criteria for the same curve -- e.g. we divided it by the normal-state conductance (vertically shifted until its tails coincide with those of the curve to be normalized) or by a polynomial fit of its high-energy tails. For any given contact, the spectra obtained by different normalization criteria were fitted independently; then, the amplitudes of the gaps were averaged and the spread of gap values obtained in all the fits was used to express the uncertainty on the gap values. In this way the uncertainty associated to the normalization is treated on the same footing as the other sources of uncertainty.
\begin{figure}
\begin{center}
\includegraphics[width=0.8\columnwidth]{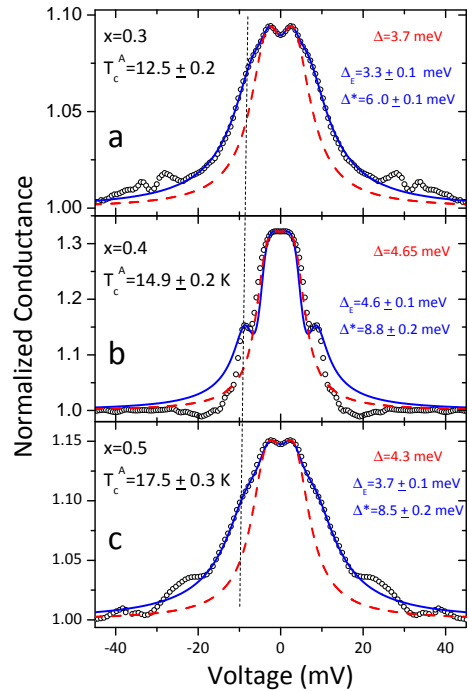}
\end{center}
\vspace{-5mm}\caption{Three examples of normalized PCARS spectra (symbols) in films with different Se content, i.e. $x=0.3$, 0.4 and 0.5 (from top to bottom). Red dashed lines: fit of the curves with the single-gap BTK model; the amplitude of the gap $\Delta$ is reported in the top right label of each plot. Solid blue lines: fit of the spectra with the two-``gap'' BTK model (see text for details). The amplitude of the gap $\Delta_E$ and of the energy scale $\Delta^*$ (that actually corresponds to the EBI) are reported in the bottom right label of each plot. }\label{fig:PCARS}
\end{figure}

Figure \ref{fig:PCARS} shows three examples, one for each doping content, of normalized conductance curves (symbols). Only the result of \emph{one} of the possible normalizations is shown for clarity. {\color{blue}{In particular, here we divided each raw spectrum by a quartic curve that fits its high-energy tails.}}
All the spectra feature a pair of conductance maxima (or a smooth plateau) at an energy of the order of 3 meV, which are the typical features associated to a nodeless superconducting gap. Additional structures are present as well, that can take the form of smooth, but well pronounced shoulders (as in panels a and c), or even small peaks (as in panel b). The shape of these structures is, in all cases, perfectly compatible with that expected for the features associated to a larger superconducting gap and indeed curves very similar to these have been actually measured in various multiband compounds, including hole- and electron-doped Ba-122 (see \cite{daghero11} and references therein). The problem here is that the energy of these structures is of the order of 8-9 meV. For example, in the spectrum of figure \ref{fig:PCARS}b where the peaks are easy to identify, their energy is $\pm 8.8$~meV. A gap of this amplitude would result in a gap ratio $2\Delta/k_B T_c \approx 13$ which looks absolutely unreasonable, even for iron-based compounds where values of the order of 9 have been sometimes found \cite{daghero11}. It is thus rather likely that these structures are the signature of another energy scale, which is \emph{not} a superconducting gap but pertains to the superconducting state as well, because the structures disappear at the critical temperature of the contact. It is thus practically impossible to ``isolate'' the spectral information about the energy gap and to get rid of these structures \footnote{The application of a magnetic field would allow solving the problem if these structures were completely suppressed by a given magnetic field, leaving the Andreev signal almost unchanged. Unfortunately, instead, the shoulders and the Andreev signal are both gradually suppressed by the field.}, which moreover have a very high spectral weight.

In these conditions, it is not even clear whether a single-gap fit that completely disregards them provides a reliable amplitude of the superconducting gap. Dashed lines in Figure \ref{fig:PCARS} represent the results of fitting the spectra with the standard 2D-Blonder-Tinkham-Klapwijk (BTK) model for a \emph{single}-gap superconductor \cite{kashiwaya00}. This model contains three parameters, i.e. the gap amplitude $\Delta$, the effective broadening $\Gamma$, and a dimensionless parameter $Z$ that accounts for both the height of the potential barrier at the interface and the mismatch of the Fermi velocities between the normal metal and the superconductor. The amplitudes of the gap obtained from the fit are reported in the top right label in each panel of figure \ref{fig:PCARS}. Even though they certainly provide the order of magnitude of the energy gap, these values suffer from the uncertainty on what has to be considered the real amplitude of the Andreev signal associated to the superconducting gap, because of the contemporary presence of the higher-energy structures.

Keeping in mind that the larger energy scale is likely not to be a gap, let us try to fit the curves by using a two-band BTK model, \emph{as if} both the structures around 3 meV and 9 meV were due to superconducting gaps. Let us call $\Delta_{E}$ the amplitude of the true superconducting gap and $\Delta^*$ the energy scale of the additional structures. The reason to do so will be clear in the following.
Let us then use an effective \emph{two}-``gap'' 2D BTK model in which the normalized conductance is expressed as a weighted sum of two contributions, i.e. $G= w_E \sigma_E + w^{*} \sigma^*$ \cite{daghero10}. The fitting function thus contains three parameters for each contribution, i.e. the energy scale ($\Delta_E$ or $\Delta^*$), the effective broadening ($\Gamma_E$ or $\Gamma^*$), and the barrier parameter ($Z_E$ or $Z^*$). Also the weight $w_E$ (or $w^*=1-w_E$) is a free parameter. The number of parameters makes the fit be non-univocal, in the sense that there is actually a range of best-fitting parameters for a single experimental curve.
The results of the two-component fit are reported in figure \ref{fig:PCARS} as solid blue lines. This model is surprisingly effective in reproducing all the main features of the curves. The values of $\Delta_E$ and $\Delta^*$, with the relevant uncertainty, are reported in the labels. Note that $\Delta_E$ has a small uncertainty because it is associated to rather sharp conductance maxima whose position is not affected by the choice of the normalization criterion. Moreover, the amplitude of $\Delta_E$ is smaller (especially in panels a and c) than the value obtained by means of the single-gap fit. Finally, it is possible to show that there is no correlation between the values of $\Delta_E$ and $\Delta^*$ and the contact resistance, which indicates that the features  we have fitted as gaps are not artifacts due, for example, to the non-spectroscopic nature of the contacts \cite{naidyuklibro,chen10b}. On the contrary, the values of $\Delta_E$ scale rather well with the local critical temperature $T\ped{c}^A$ giving a constant gap ratio $2\Delta_E/k_BT_c \simeq 5.5$, which is well above the BCS weak-coupling limit but not abnormal in Fe-based compounds. The values of $\Delta^*$ are more scattered, but their overall trend as a function of $T\ped{c}^A$ can be approximately expressed as $2 \Delta^* /k_{B}T_{c} \approx 11.5$.

Two points must then be clarified: i) what is $\Delta^*$; ii) which of the two fitting procedures (with a single-gap or a two-``gap'' model) gives the correct value of the superconducting gap $\Delta_E$.

As for the first point, a possible cause of the structures at $\Delta^*$ could be the strong electron-boson interaction (EBI). As a matter of fact, we have shown in Co-doped Ba-122 \cite{tortello10} and in F-doped Sm-1111 \cite{daghero11} that EBI structures are indeed observable by point-contact spectroscopy not only in the tunneling regime but also in the Andreev-reflection regime. We have also shown that these structures can be accounted for rather well by inserting into the BTK model the energy-dependent order parameters calculated, within the Eliashberg theory, by using a Lorentzian electron-boson spectrum $\alpha^2 F(\Omega)$ peaked at the energy of the spin resonance $\Omega_0$ (measured by inelastic neutron scattering experiments \cite{qiu09}) that scales with $T\ped{c}$ according to the empirical law $\Omega_0 = 4.65 k\ped{B}T\ped{c}$ \cite{paglione10}.  As shown elsewhere \cite{daghero11}, the peak in the $\alpha^2 F(\Omega)$ results in a peak in the sign-changed derivative of the conductance, $-d^2 I /dV^2$ vs. $V$, that approximately occurs at $\Omega_0 + \Delta_{max}$.

\begin{figure}
\begin{center}
\includegraphics[width=\columnwidth]{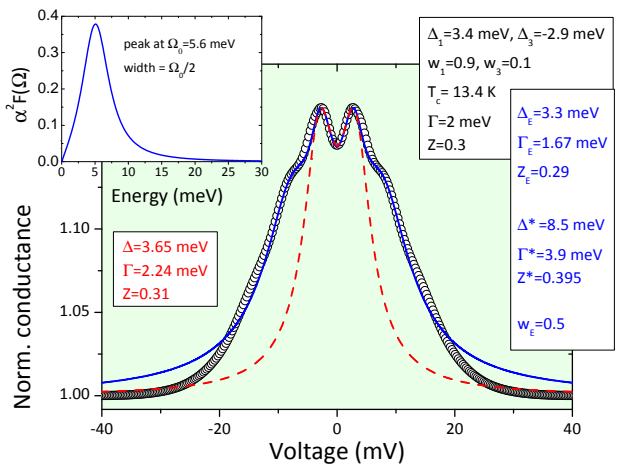}
\end{center}
\vspace{-3mm}\caption{Symbols: theoretical PCARS spectrum calculated within the two-band 2D BTK model by using the energy-dependent order parameters $\Delta_1$ and $\Delta_3$ obtained from the Eliashberg model. The parameters of the BTK model are reported in the top right label, including the amplitude of the gaps whose different sign is due to the $s \pm$ symmetry. The theoretical spectrum presents, in addition to the conductance maxima due to the gaps (almost identical in amplitude) clear shoulders due to the EBI. Solid blue line: fit of the theoretical PCARS spectrum with a two-band 2D BTK model. The fit is made by treating the EBI shoulders as if they were due to a large gap $\Delta^*$. The best-fit parameters are listed in the bottom right label. Red dashed line: fit of the theoretical PCARS spectrum with a single-band 2D BTK model. The best-fitting parameters are reported on the left label. Inset: the shape of the electron-boson spectrum (Eliashberg function) used in the calculations.}\label{fig:eliashberg}
\end{figure}

Going back to Fe(Te,Se), let us assume that $\Delta_E$ is the true energy gap, and check whether the coupling of electrons with spin fluctuations can give rise to higher-bias additional structures similar to those observed experimentally. For example, let us focus on $\mathrm{Fe(Te_{1-x}Se_x)}$ with a critical temperature of about 14 K and $\Delta_E= \frac{5.5}{2} k_B T_c = 3.3 $ meV. Recent angle-resolved photoemission spectroscopy (ARPES) measurements have given direct evidence of 3 bands crossing the Fermi level in FeTe$_{0.55}$Se$_{0.45}$ with $T_c = 14.5$ K, i.e. two holelike bands at the $\Gamma$ point of the Brillouin zone and one electronlike band at the $M$ point \cite{miao12}. DFT calculations \cite{zeng10} give evidence of two holelike FS sheets (almost perfectly cylindrical) and two electronlike FS sheets, of which the inner one is almost cylindrical while the outer one displays a strong warping. To build up our effective three-band Eliashberg model, we used two effective holelike Fermi surfaces (labeled as 1 and 2 in the following) at $\Gamma$ and one electronlike FS (labeled as 3) at M. We estimated the ratio of the density of states at the Fermi level from DFT calculations, obtaining $N_1/N_3 = 0.46$ and $N_2/N_3 =1$, and we made as usual the following assumptions (for details see ref.\cite{ummarino09,ummarino11}): i) the contribution of phonons to the coupling is negligible; ii) the interband coupling is mediated by spin fluctuations, while the intraband coupling is negligible; iii) the Eliashberg function $\alpha^2F(\Omega)$ has the same shape for all coupling channels, but its height is modulated by the corresponding coupling constant (i.e. $\lambda_{12}, \lambda_{13}, \lambda_{23}$); iv) the coupling between the two holelike bands is negligible, so $\lambda_{12} \simeq 0$; v) the $\alpha^2F(\Omega)$ is a Lorentzian curve peaked at the energy of the spin resonance, $\Omega_0=4.65 k_{B}T_{c}$ \cite{paglione10}, and its half-width at half maximum is $\Omega_0/2$ (see left inset to figure  \ref{fig:eliashberg}); vi) the Coulomb pseudopotential is negligible (and thus we take $\mu^*_{i,j}=0$ for any $i,j$.

In the end, the model contains only two free parameters, $\lambda_{13}$ and $\lambda_{23}$, that can be adjusted to reproduce the values of the gaps. Since the position of the EBI structures depends on the energy of the mediating boson and on the largest superconducting gap, and since we just want to check whether these structures can be mistaken by a gap, we use the simplest possible assumption, i.e. that two gaps exist of similar amplitude. Once $\lambda_{13}$ and $\lambda_{23}$ are determined, the critical temperature is calculated with no additional adjustment of the parameters. We found that the experimental situation ($T_c$ of about 14 K, and one single gap amplitude of about $3.3$ meV) can be obtained by using $\lambda_{23}=0$ and $\lambda_{13}=4.1$ (this value looks large but corresponds to a total coupling constant $\lambda_{tot}=1.56$).  In particular, we got $T_c=13.4$ K, $\Delta_1=3.4$ meV, $\Delta_2=0$, $\Delta_3=-2.9$ meV. The two non-zero gaps are very similar in amplitude but differ in sign because of the $s\pm$ symmetry. The problem has thus been reduced to a two-band one and therefore the energy-dependent order parameters can be inserted into the two-band 2D BTK model to calculate the normalized conductance. To keep the same labels for the bands, the conductance can be conveniently expressed as $G(V)=w_1 \sigma_1(V) + (1-w_1)\sigma_3(V)$.  The resulting curve always shows maxima due to the gaps at about 3 meV, plus additional shoulders or even small peaks due to the EBI. A curve qualitatively similar, in amplitude and shape, to the experimental ones is shown in figure \ref{fig:eliashberg} (symbols). It was obtained by choosing $w_1=0.9$, $Z_1=Z_3=0.3$ and $\Gamma_1=\Gamma_3=2$ meV. For a further check, we can try to fit this curve with the two-band 2D BTK model with constant (BCS) energy gap, thus doing exactly what we did with the experimental spectra, and treating the EBI structures as if they were due to a larger gap. The result of the fit is shown by a solid blue line in figure \ref{fig:eliashberg}, and the corresponding parameters are listed in the label. Note that the fit gives $\Delta_E=3.3$ meV, in very good agreement with the original amplitude of the gap with which the curve was generated, and $\Delta^*=8.5$ meV, which is perfectly compatible with the values of $\Delta^*$ obtained from the two-gap fit of the experimental PCARS spectra at $T\ped{c}^A \simeq 14$ K. Instead, the single-gap fit of the conductance curve (dashed red line in figure \ref{fig:eliashberg}) would give a gap amplitude $\Delta=3.65$ meV which is slightly overestimated.

The above discussion proves that: i) the high-energy structures observed in the PCARS spectra are very likely to be due to the EBI; ii) $\Delta^*$ is not a gap, but rather the energy at which the EBI manifests itself in the conductance; {\color{blue}{iii}}) the amplitude of the superconducting gap $\Delta_E$ is better reproduced by the two-gap BTK fit than by the single-gap fit (see for instance figure \ref{fig:PCARS}).

\begin{figure}
\begin{center}
\includegraphics[width=0.8\columnwidth]{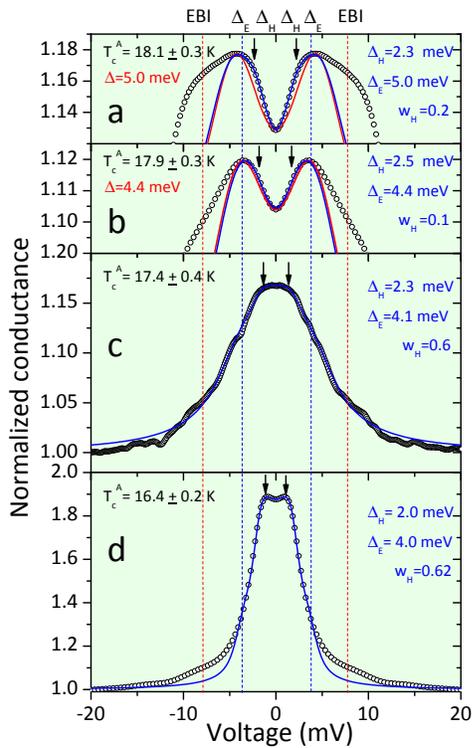}
\end{center}
\vspace{-5mm}
\caption{{\color{blue}{Some}} examples of PCARS curves giving evidence for the smaller gap $\Delta_H$. In all cases the gap amplitude obtained from the fit are indicated in the labels. {\color{blue}{(a,b) Detail of two spectra measured in Fe(Te,Se) thin films with $x=0.5$ that show a small excess conductance inside the maxima associated to $\Delta_E$ that cannot be fitted by a single-gap model (red line) and is instead perfectly captured by a two-gap model (blue line) with $\Delta_E$ and the inner gap $\Delta_H$. (c) one of the few curves in the $x=0.5$ films where the EBI structures are less pronounced and a two-band fit with the gaps $\Delta_H$ and $\Delta_E$ is able to reproduce most of the curve. (d) A PCARS spectrum taken in a single crystal, in which the superconducting signal is almost ideal, with the relevant fit (blue line). Vertical dashed lines approximately indicate the position of the features associated to the EBI and to the gap $\Delta_E$. Arrows indicate the structures associated to the smaller gap $\Delta_H$.}}} \label{fig:smallgap}
\end{figure}

\begin{figure}
\begin{center}
\includegraphics[width=\columnwidth]{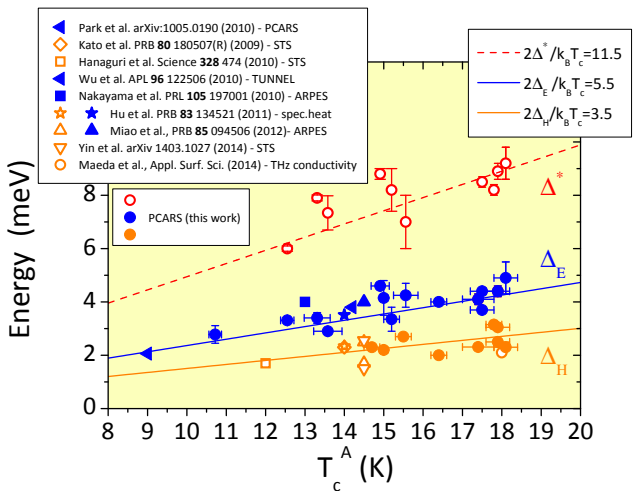}
\end{center}
\vspace{-8mm}
\caption{Gap amplitudes $\Delta_H$ (orange symbols) and $\Delta_E$ (blue symbols) and EBI energy $\Delta^*$ (red open symbols) as a function of the local critical temperature $T\ped{c}^A$. Data from literature are added for comparison; in these cases, the critical temperature is the one declared in the original papers. The data come from PCARS \cite{park10}, STS \cite{kato09,hanaguri10,yin14}, ARPES \cite{nakayama10,miao12}, tunnel spectroscopy \cite{wu10}, specific heat \cite{hu11b} and THz spectroscopy \cite{maeda14}.} \label{fig:gaps_vs_Tc2}
\end{figure}

We can now make a step forward in order to understand whether $\Delta_E$ is the only superconducting gap detected by PCARS. Interestingly, this gap is in very good agreement with the gap that has been recently measured by ARPES on the electronlike FS $\gamma$ \cite{miao12}. Actually, in the films with $x=0.5$ there are some experimental facts that suggest that a second, smaller gap might be present as well, even though with a small weight \footnote{In the films with $x=0.3$ and $x=0.4$ instead there is no such clear evidence, possibly because the second gap is too small to be detected.}. In some curves, the single-gap fit or even the two-gap fit (with $\Delta_E$ and $\Delta^*$) are not completely satisfactory in the low-energy region {\color{blue}{ $eV < \Delta_E$, where a small excess conductance exists (see panels a and b in fig. \ref{fig:smallgap})}}. This small discrepancy can be removed if the low-energy part of the curves is fitted (disregarding the EBI structures) by means of the two-band BTK model and, in addition to $\Delta_E$, a second smaller gap $\Delta_H$ is considered (blue solid lines in figure \ref{fig:smallgap}a and b). The fit is obtained here with $\Delta_E=5.0 \pm 0.3$ meV and $\Delta_H=2.3 \pm 0.1$ meV (panel a) and $\Delta_E=4.4 \pm 0.2$ meV and $\Delta_H=2.5 \pm 0.1$ meV {\color{blue}{ (panel b)}}. A more convincing evidence for the existence of the small gap $\Delta_H$ comes from some PCARS curves in which the EBI shoulders are hardly detectable (for unknown reasons) and the Andreev-reflection structures are not, or poorly,  disturbed by them. An example is given in  figure  {\color{blue}{\ref{fig:smallgap}c}}. The fit of this curve again gives $\Delta_H=2.3 \pm 0.1$ meV and $\Delta_E=4.1 \pm 0.2$ meV. Finally, the most striking proof of the fact that an additional smaller gap exists comes from PCARS measurements carried out, with the same technique and in the same configuration ($c$-axis injection) in single crystals of $\mathrm{Fe(Te_{0.8}Se_{0.2})}$. An example of these curves is shown in figure  {\color{blue}{\ref{fig:smallgap}d. Note that the local critical temperature of this contact is higher than $T_c^{90}$ of the crystal (see figure \ref{fig:resistivity}b). This anomaly has been already observed in $\mathrm{Fe(Te_{0.55}Se_{0.45})}$ crystals \cite{park10} and might be due to a different local concentration of excess Fe. In our case, the point contacts were made on a fresh surface exposed by cleaving the crystal while the contacts for the resistance measurements were placed on the original surface. This, taking into account the process of outward migration of interstitial Fe atoms induced by annealing, probably explains the discrepancy in the critical temperatures.}} The superconducting signal is extremely high here, close to the theoretical limit of 2, which means the contact is nearly ideal. The experimental curve shows two clear maxima corresponding to $\Delta_H$, a change in slope at an energy corresponding to $\Delta_E$ and wide, but much lower, EBI shoulders at higher energy. The two-band fit of the curve (neglecting the EBI structures) gives $\Delta_H = 2.00 \pm 0.05$ meV and $\Delta_E=4.0 \pm 0.1$ meV.
Let us now consider together all the information on the three energy scales obtained so far, i.e.: i) the results of the two-band BTK fit of the conductance curves that give evidence of $\Delta_E$ and of the EBI structures at $\Delta^*$ and not of $\Delta_H$. All the PCARS spectra in the films with $x=0.3$ and $x=0.4$, but also some of the spectra in the $x=0.5$ film, are of this kind (see figure \ref{fig:PCARS}); ii) the results of the fit of all the spectra that also provide evidence for the smaller gap $\Delta_H$, like those in figure \ref{fig:smallgap} and others obtained in similar situations.
All the available values of $\Delta_H$, $\Delta_E$ and $\Delta^*$ are plotted in figure \ref{fig:gaps_vs_Tc2} as a function of the local critical temperature of the contacts, $T\ped{c}^{A}$. A clear picture emerges in which all the three energy scales depend on the critical temperature in a linear way. The gaps scale rather well with the critical temperature according to a constant gap ratio, i.e. $2\Delta_H/k_BT\ped{c}^A=3.5$ for $\Delta_H$ and $2\Delta_E/k_BT\ped{c}^A=5.5$ for $\Delta_E$. For the energy $\Delta^*$, the values are more scattered but approximately fall on the line $2\Delta^*/k_BT\ped{c}^A \simeq 11.5 $.
The same figure also reports the results of many measurements of the energy gaps in $\mathrm{FeTe_{1-x}Se_{x}}$ taken from literature; in these cases, the critical temperature is that declared in the original paper. The agreement with our results is excellent; actually, the systematic investigation of the energy gaps as a function of $T\ped{c}^A$ (and thus of the doping) performed here for the first time allows understanding and explaining within a single simple picture the apparent scattering of gap data present in literature. It is worthwhile to remark that our data perfectly agree with recent results of ARPES \cite{miao12} and STS \cite{yin14} as well as with the results of THz conductivity carried out in the same films at $x=0.5$ \cite{maeda14}. With respect to the ARPES result, we can conclude that PCARS can detect very easily the gap $\Delta_E$ associated to the electronlike FS, and with more difficulty a gap $\Delta_H$ that is probably the average of the gaps residing on the two holelike pockets, and that are too close to be resolved by PCARS. The easier detection of $\Delta_E$ might be explained by the fact that the electronlike FS is more 3D than the holelike ones and this makes its weight for $c$-axis conduction be greater. This explanation however conflicts with the fact that STS measurements \cite{yin14} performed with $I \parallel c$ axis, exactly as in our case, detect the two holelike gaps and not the electronlike one. The reason for this discrepancy may be due to the higher directionality of tunneling spectroscopy (so that it  mainly probes states with small in-plane momentum component and thus is more sensitive to the region around $\Gamma$ \cite{golubov13}) with respect to Andreev-reflection spectroscopy. Indeed, in the pure Andreev-reflection regime ($Z=0$) the normal-state probability of electron injection is isotropic (i.e. identical for all directions in the whole half-space). In these conditions, the ``weight'' of each FS sheet in the spectra is proportional to the area of its projection on a plane perpendicular to the direction of current injection \cite{daghero13}, which means that the FS sheets with enhanced 3D character should dominate the conductance for $I\parallel c$. In the pure tunneling regime ($Z=\infty$), instead, the probability of electron injection strongly decreases on going away from the normal direction, and this makes the FS sheets with small transverse component of $k$ be dominant in the spectrum.

It is worth noting that we have always fitted the experimental curves by using \emph{isotropic} gaps. This is due to the fact that the PCARS spectra do not show any clear evidence of nodes in the order parameters, although this does not exclude the presence of local gap minima in some of the Fermi surfaces. This said, in $c$-axis contacts we have found no clear hints in favor of the fourfold gap anisotropy observed by directional specific-heat experiments \cite{zeng10}. It is true, however, that also ARPES \cite{miao12} and STS \cite{yin14} results are compatible with isotropic gaps. Maybe a possible explanation of this disagreement is that given in ref. \cite{miao12} where the anisotropy observed by directional specific heat is ascribed to the anisotropy of the Fermi surface with respect to the $\Gamma$ point, rather than to the anisotropy of the gap on a single specific FS sheet. To investigate this point in greater detail, further PCARS measurements in single crystals (with the current injected along the $ab$ planes) are underway and will be the subject of a forthcoming paper.
\bigskip

\section{Conclusions}
We have presented the results of the first extensive study of the superconducting gaps in $\mathrm{Fe(Te_{1-x}Se_{x})}$ with various Se contents, i.e $x=0.3$, $x=0.4$ and $x=0.5$. The gaps have been determined by means of point-contact Andreev-reflection spectroscopy in epitaxial films grown by PLD on single-crystalline CaF$_2$ substrates. The PCARS spectra generally show clear symmetric maxima associated to a superconducting gap of amplitude $\Delta_E \simeq 2.75 k_{B}T_{c}$ and additional, very clear shoulders that can be mistaken for a gap of amplitude $\Delta^* \approx 6 k_{B}T_{c}$ but are very probably the signature of the strong coupling of electrons with a bosonic mode peaked at an energy $\Omega_0$ -- that roughly obeys the empirical rule $\Omega_0=4.65 k_{B}T_c$ demonstrated for the spin resonance energy. A comparison with the results of ARPES suggests that the gap $\Delta_E$ might be located on the electronlike FS sheet $\gamma$.
A careful analysis of the low-energy region of some spectra taken in the $x=0.5$ films suggests the existence of a (hardly detectable) smaller gap $\Delta_H \simeq 1.75 _B T_c $. More reliable evidences of this smaller gap come from the fit of a few spectra where the EBI structures are partly suppressed, and even more from PCARS measurements in single crystals --  which clearly show both the gaps $\Delta_E$ and $\Delta_H$ \emph{and} the EBI structures. Once plotted as a function of the local critical temperature, $\Delta_E$ and $\Delta_H$ provide a unifying framework in which all the data reported in literature (and characterized up to now by an apparently unreasonable spread) perfectly fit. In particular, the smaller gap $\Delta_H$ turns out to be the ``average'' of the two gaps residing on the holelike FS sheets recently identified by ARPES and STS.

This work was done under the Collaborative EU-Japan Project ``IRON SEA'' (NMP3-SL-2011-283141).

\end{document}